\documentclass{elsart}
\usepackage{epsf,psfig}
\begin{document}

\begin{frontmatter}

\title{Pairing correlations and transitions in nuclear systems}

\author[belgrade]{A.~Beli\'c},
\author[ornl]{D.~J.~Dean}, and
\author[oslo]{M.~Hjorth-Jensen}
\address[belgrade]{Institute of Physics, P.O.B. 57, Belgrade 11001, Serbia and Montenegro}
\address[ornl]{Physics Division, Oak Ridge National Laboratory,
P.O. Box 2008, Oak Ridge, TN 37831-6373, USA}
\address[oslo]{Department of Physics and Center of Mathematics for Applications,
         University of Oslo, N-0316 Oslo, Norway}

\begin{abstract}

We discuss several pairing-related
phenomena in nuclear systems, ranging from superfluidity
in neutron stars to the gradual breaking of pairs in finite nuclei.
We  describe recent experimental evidence that
points to a relation between pairing and
phase transitions (or transformations) in finite nuclear
systems. A simple pairing interaction
model is used in order to study and classify an
eventual pairing phase transition in finite
fermionic systems such as nuclei.
We show that systems with as few as $\sim 10-16$ fermions
can exhibit clear features reminiscent of a phase transition.
\end{abstract}

\end{frontmatter}

\section{Introduction}

The standard BCS theory has been widely used to describe 
systems with pairing correlations and phase transitions to a
superconducting phase for large systems, 
from the solid state to nuclear physics,
with neutron stars as perhaps the largest object in the universe
exhibiting  superfluidity in its interior.  
An eventual superfluid phase in a neutron star will condition 
the neutrino emission and thereby the cooling history of such a
star, in addition to inducing mechanisms such as 
sudden spin ups in the rotational
period of the star; see, for example, Ref.~\cite{hh2000,dh2003} 
for recent reviews. 
For an infinite system, such as a neutron star, the nature
of the pairing phase transition is well established as second order.
 
When a system of correlated fermions such as electrons or
nucleons is sufficiently small,
the fermionic spectrum becomes
discrete. If the spacing approaches the size of the pairing gap,
superconductivity is expected to break down \cite{anderson59};
however, recent experiments on superconducting ultrasmall aluminum 
grains by Tinkham
{\em et al.} \cite{tinkham9598} revealed the existence of a
spectroscopic gap larger than the average electronic level density.
This feature was interpreted as a reminiscence of superconductivity
and renewed the interest \cite{delft98,mastellone98,sierra99,delft2000}
in studies  of what is the lower size limit for superconductivity.

Other finite fermionic systems such as   
nuclei are expected to
exhibit a variety of interesting phase-transition like phenomena, like
the disappearence of pairing at a critical temperature $T_c\approx
0.5-1$ MeV or the nuclear shape transitions of deformed nuclei
associated with the melting of shell effects at  
$T_c\approx 1-12$ MeV.  
Pairing correlations are expected 
to play an essential role in nuclear systems, ranging 
from the binding energy, excitation spectrum and odd-even effects
in finite nuclei to superfluidity in the interior of neutron stars.
In recent theoretical and experimental studies \cite{yoram2000,andreas2000}
of thermodynamical properties of finite nuclei, the heat capacity 
has been found to exhibit a non-vanishing bump at temperatures proportional to
half the pairing gap. These bumps were interpreted as  signs of the 
quenching of pair correlations, representing in turn features
of the pairing transition for an infinitely large system. 
In the study of eventual 
transitions in e.g., nuclear physics, it is important to know  whether
a given transition really is of first order, discontinuous, 
or if there is a continuous change
in a physical quantity like the mean energy, as in phase transitions
of second order. If one works in the canonical or grand canonical
ensembles, for finite systems
it is rather difficult to decide on the order of the phase
transition. This is due to the fact that in ensembles like the canonical,
any anomaly is smeared over a temperature range of $1/N$, $N$ being the
number of particles. In the analysis of  finite systems, both a 
$\delta$-function peak and a power law singularity sharpen as the
number of particles is increased, making it difficult to distinguish
between the two cases, see, for example, Ref.~\cite{huller}. 
In addition, first order phase transitions in finite systems 
have recently been inferred, theoretically and experimentally,  from 
observed negative heat capacities that 
are associated with anomalous convex intruders in the entropy
versus energy curves, resulting in backbendings in the caloric 
curves; see, for example, 
Refs.~\cite{andreas2000,huller,gross,schmidt01,agostino00,gc00}.
Negative heat capacities are often claimed to appear only in calculations
done in the microcanonical ensemble and are thought to vanish 
in the canonical or grand-canonical ensembles. 

In this work we give first a brief review in Sec.~\ref{sec:sec2} 
of pairing features in
infinite neutron matter. In Sec.~\ref{sec:sec3} we discuss experimental results
indicating the gradual breaking of pairs in nuclei. A simple pairing model
is in turn used in Sec.~\ref{sec:sec4} 
to show the similarities between the experimental results and
the gradual breaking of pairs.
Concluding remarks are presented in Sec.~\ref{sec:sec5}.

\section{Pairing in infinite neutron matter}\label{sec:sec2}
The presence of neutron superfluidity in 
the crust and the inner part 
of neutron stars 
are considered well established 
in the physics of these compact stellar objects. 
In the low density outer part of a neutron star, 
the neutron superfluidity is expected 
mainly in the attractive $^1S_0$ channel. 
At higher density, the nuclei in the crust dissolve, and one 
expects a region consisting of a quantum liquid of neutrons and 
protons in beta equilibrium. 
The proton contaminant should be superfluid 
in the $^1S_0$ channel, while neutron superfluidity is expected to  
occur mainly in the coupled $^3P_2$-$^3F_2$ two-neutron channel. 
In the core of the star any superfluid 
phase should finally disappear.
 
The presence of two different superfluid regimes 
is suggested by the known trend of the 
nucleon-nucleon (NN) phase shifts 
in each scattering channel. 
In both the $^1S_0$ and $^3P_2$-$^3F_2$ channels the
phase shifts indicate that the NN interaction is attractive. 
In particular for the $^1S_0$ channel, the occurrence of 
the well known virtual state in the neutron-neutron channel
strongly suggests the possibility of a 
pairing condensate at low density, 
while for the $^3P_2$-$^3F_2$ channel the 
interaction becomes strongly attractive only
at higher energy, which therefore suggests a possible 
pairing condensate
in this channel at higher densities. 
In recent years the BCS gap equation
has been solved with realistic interactions, 
and the results confirm
these expectations. 

The $^1S_0$ neutron superfluid is relevant for phenomena
that can occur in the inner crust of neutron stars, like the 
formation of glitches, which may to be related to vortex pinning  
of the superfluid phase in the solid crust \cite{glitch}. 
The results of different groups are in close agreement
on the $^1S_0$ pairing gap values and on 
its density dependence, which
shows a peak value of about 3 MeV at a Fermi momentum close to
$k_F \approx 0.8\; {\rm fm}^{-1}$ \cite{bcll90,kkc96,eh98,sclbl96}. 
All these calculations adopt the bare
NN interaction or effective interactions without screening 
corrections as the pairing force. It has been pointed out
that the screening by the medium of the interaction 
could strongly reduce
the pairing strength in this channel \cite{sclbl96,chen86,ains89}. 
However, the issue of the 
many-body calculation of the pairing 
effective interaction is a complex
one and still far from a satisfactory solution.

The precise knowledge of the $^3P_2$-$^3F_2$ pairing gap is of 
paramount relevance for, e.g.,  the cooling of neutron stars, 
and different values correspond to drastically
different scenarios for the cooling process.
Generally, the gap suppresses the cooling by a factor
$\sim\exp(-\Delta/T)$ (where $\Delta$ is the energy gap)
which is severe for
temperatures well below the gap energy.

For $\beta$-stable matter in equilibrium, the neutron
$^1S_0$ pairing gap appears at densities corresponding to the crust 
of the star. It is generally believed that it is the
proton contaminant and its $^1S_0$ pairing gap which dominates
in the region from the inner crust to the densities 2-3 times nuclear matter
saturation density, together with the   $^3P_2$ gap. The general picture
can be summarized as follows:
\begin{itemize}
\item
The $^1S_0$ proton gap in $\beta$-stable matter
            is $ \le 1$ MeV, and if polarization
            effects were taken into account \cite{sclbl96},
            it could be further reduced by a factor 2-3.
\item
The $^3P_2$ gap is also small, of the order
            of $\sim 0.1$ MeV in $\beta$-stable matter.
            If relativistic effects are taken into account,
            it is almost vanishing. However, there is
            quite some uncertainty with the value for this
            pairing gap for densities above $\sim 0.3$
            fm$^{-3}$ due to the fact that the NN interactions
            are not fitted for the corresponding lab energies. 
\item
Higher partial waves give essentially vanishing
            pairing gaps in $\beta$-stable matter.
\end{itemize}
Thus, the $^1S_0$ and $^3P_2$ partial waves are crucial for our
understanding of superfluidity in neutron star matter. However,
hyperons such as $\Sigma^{-1}$ and $\Lambda$ may be present at twice or more
nuclear matter saturation energy. 
There are indications that the $\Lambda\Lambda$ interaction is too weak
to support a $\Lambda$ gap, while $\Delta_{\Sigma^{-1}}\sim 10$ MeV. 
Recent cooling simulations seems to indicate
that available observations of thermal emissions from pulsars can aid
in constraining hyperon gaps. However, all these calculations suffer from the
fact that the microscopic inputs, pairing gaps, composition of matter, emissivity rates, etc.
are not computed at the same many-body theoretical level. This leaves a 
considerable uncertainty. 

We have not mentioned recent developments beyond the BCS approach,
nor have we discussed results for proton-neutron pairing in symmetric
or asymmetric matter. Such topics are addressed in the recent works
of Lombardo, Schulze and collaborators, see e.g., Refs.\ 
\cite{dh2003,ls2000,ls2001} and references therein.

\section{Thermodynamic properties of nuclei and pairing}\label{sec:sec3}

The thermodynamical properties of nuclei deviate from infinite systems, 
although the spectroscopy of finite nuclei and especially many isotopes, 
are dominated by the same partial waves which are important 
in neutron star matter, see ref.~\cite{dh2003}.

While the quenching of pairing in superconductors is well described as a 
function of temperature, the nucleus represents a finite many body system 
characterized by large fluctuations in the thermodynamic observables. A 
long-standing problem in experimental nuclear physics has been to observe the 
transition from strongly paired states, at around $T=0$, to unpaired states at 
higher temperatures. 

In nuclear theory, the pairing gap parameter $\Delta$ can be studied as 
function of temperature using the BCS gap equations \cite{SY63,Go81}. From this
simple model the gap decreases monotonically to zero at a critical temperature 
of $T_c\sim 0.5\,\Delta$. However, if particle number is projected out 
\cite{FS76,DK95}, the decrease is significantly delayed. The predicted decrease
of pair correlations takes place over several MeV of excitation energy 
\cite{DK95}. Recently \cite{andreas2000}, structures in the level 
densities in the 1--7~MeV region were reported, structures which 
probably are due to the breaking of 
nucleon pairs and a gradual decrease of pair correlations. 

Experimental data on the quenching of pair correlations are important as a 
test for nuclear theories. Within finite temperature BCS and RPA models, level 
density and specific heat are calculated for e.g., $^{58}$Ni \cite{Ng90}; 
within the shell model Monte Carlo method (SMMC) \cite{LJ93,KD97} one is now 
able to estimate level densities \cite{Or97} in heavy nuclei \cite{WK98} up to 
high excitation energies. 
Here we report on the observation of the gradual 
transition from strongly paired states to unpaired states in rare earth nuclei 
at low spin. The canonical heat capacity is used as a thermometer. Since only 
particles at the Fermi surface contribute to this quantity, it is very 
sensitive to phase transitions. It has been demonstrated from SMMC calculations
in the Fe region \cite{RH98,AL99}, that breaking of only one nucleon pair 
increases the heat capacity significantly. 

The experiments were carried out with 45~MeV $^3$He projectiles from the MC-35 
cyclotron at the University of Oslo. In that experiment, one could
extract level densities and $\gamma$
strength functions for the $^{161,162}$Dy and $^{171,172}$Yb nuclei. The data 
for the even nuclei are published recently \cite{andreas2000}.

The partition function in the canonical ensemble 
$Z(T)=\sum_{n=0}^\infty\rho(E_n)e^{-E_n/T} $
is determined by the measured level density of accessible states $\rho(E_n)$ in
the present nuclear reaction. Strictly, the sum should run from zero to 
infinity. Here  we calculate $Z$ for temperatures up to $T=1$~MeV. 
However, the experimental level
densities only cover the excitation region up 
close to the neutron binding energy of about 6 and 8~MeV for odd and even mass 
nuclei, respectively. For higher energies it is reasonable to assume Fermi gas 
properties, since single particles are excited into the continuum region with 
high level density. Therefore, due to lack of experimental data, the level 
density is extrapolated to higher energies by the shifted Fermi gas model 
expression \cite{GC65}. 
The extraction of the microcanonical heat capacity $C_V(E)$ gives large 
fluctuations which are difficult to interpret \cite{andreas2000}. Therefore, the heat 
capacity $C_V(T)$ is calculated within the canonical ensemble, where $T$ is a 
fixed input value in the theory, and a more appropriate parameter,
see e.g., Schiller {\em et al.} \cite{andreas2000} for further details.

\begin{figure}
\begin{center}
          {\epsfxsize=20pc \epsfbox{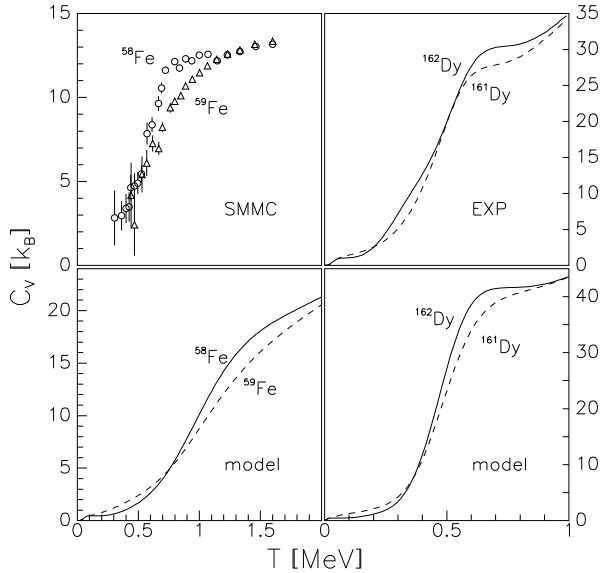}}
          \caption{Heat capacity for iron isotopes, see Ref.\ \cite{Al99}, 
and for $^{161,162}$Dy. See text for further details. \label{fig:heatcapacity}}
\end{center}
\end{figure}

\begin{figure}
\begin{center}
          {\epsfxsize=20pc \epsfbox{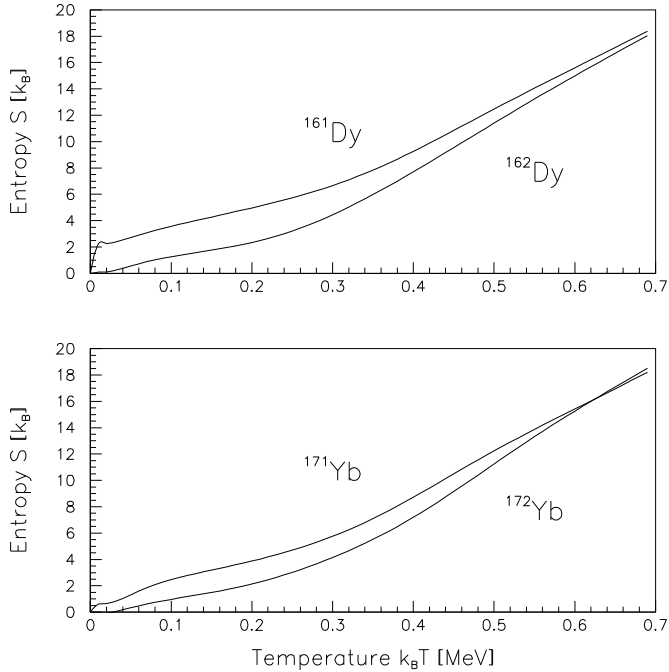}}
          \caption{Experimental entropy in the canonical ensemble for $^{161,162}$Dy and
for $^{171,172}$Yb. \label{fig:entropy}}
\end{center}
\end{figure}

The deduced heat capacities for the $^{161,162}$Dy nuclei 
are shown in Fig.~\ref{fig:heatcapacity} together with the SMMC
results of Liu and Alhassid \cite{Al99} for various iron isotopes.
The results labelled 'model' are discussed further in Refs.\ 
\cite{dh2003,andreas2000}. We note that both the theoretical and 
experimental results exhibit 
S-shaped $C_V(T)$-curves.
The S-shaped curve is interpreted as a 
fingerprint of a phase transition in a finite system from a phase with strong 
pairing correlations to a phase without such correlations. Due to the strong 
smoothing introduced by the transformation to the canonical ensemble, we do not
expect to see discrete transitions between the various quasiparticle regimes, 
but only the transition where all pairing correlations are quenched as a whole.
It is worth noticing that the S-shape is much less 
pronounced for the odd system,
again a possible indication of the importance of pairing correlations. 
This can also be seen from Fig.\ \ref{fig:entropy}, taken from Ref.\
\cite{andreas2000}.

Here we notice that the entropy of the even and odd systems merge
at a temperature $T\approx 0.5$ MeV, in close agreement with the point where
the S-shape of the heat capacity of 
the $^{161,162}$Dy nuclei appears  
in Fig.~\ref{fig:heatcapacity}. The temperature where 
the experimental entropies merge,
could in turn be interpreted as the point where other degrees of freedom
than pairing take over. A theoretical interpretation
in terms of the vanishing of pairing correlations is given 
in Refs.\ \cite{andreas2000}. 
The extraction of the microcanonical heat capacity $C_V(E)$ gives large 
fluctuations which are difficult to interpret \cite{andreas2000}. Therefore, the heat 
capacity $C_V(T)$ is calculated within the canonical ensemble, where $T$ is a 
fixed input value in the theory, and a more appropriate parameter,
see e.g., Schiller {\em et al.} \cite{andreas2000} for further details.

The deduced heat capacities for the $^{161,162}$Dy nuclei 
are shown in Fig.~\ref{fig:heatcapacity} together with the SMMC
results of Liu and Alhassid \cite{Al99} for various iron isotopes.
The results labelled 'model' are discussed further in Ref.\ 
\cite{andreas2000}. We note that both the theoretical and 
experimental results exhibit 
S-shaped $C_V(T)$-curves.
The S-shaped curve is interpreted as a 
fingerprint of a phase transition in a finite system from a phase with strong 
pairing correlations to a phase without such correlations. Due to the strong 
smoothing introduced by the transformation to the canonical ensemble, we do not
expect to see discrete transitions between the various quasiparticle regimes, 
but only the transition where all pairing correlations are quenched as a whole.
It is worth noticing that the S-shape is much less 
pronounced for the odd system,
again a possible indication of the importance of pairing correlations. 
This can also be seen from Fig.\ \ref{fig:entropy}, taken from Ref.\
\cite{andreas2000}.
Here we notice that the entropy of the even and odd systems merge
at a temperature $T\approx 0.5$ MeV, in close agreement with the point where
the S-shape of the heat capacity of 
the $^{161,162}$Dy nuclei appears  
in Fig.~\ref{fig:heatcapacity}. The temperature where 
the experimental entropies merge,
could in turn be interpreted as the point where other degrees of freedom
than pairing take over. A theoretical interpretation
in terms of the vanishing of pairing correlations is given 
in Ref.\ \cite{andreas2000} and in the next section.

\section{Simple pairing model and nature of the pairing transition}
\label{sec:sec4}

We aim here to identify the nature of the pairing
transition and give a theoretical interpretation of the results 
from the previous section.
Since we are dealing with pairing correlations, our 
Hamiltonian is 
\begin{equation}
   H=\sum_i \varepsilon_i a^{\dagger}_i a_i -G\sum_{ij>0}
           a^{\dagger}_{i}
     a^{\dagger}_{\bar{\imath}}a_{\bar{\jmath}}a_{j},
     \label{eq:pairHamiltonian}
\end{equation}
where $a^{\dagger}$ and $a$ are fermion creation and annihilation operators, 
respectively. The indices $i$ and $j$ run over the number 
of levels $L$, and the label $\bar{\imath}$ stands for a time-reversed state. 
The parameter $G$ is the strength of the pairing 
force while $\varepsilon_i$ is 
the single-particle energy of level $i$. 
We assume that the single-particle levels are equidistant with a 
fixed spacing $d$.
Moreover, in our simple model, the degeneracy of the single-particle 
levels is set to $2J+1=2$, with $J=1/2$ being the spin of the particle. 
Seniority $\bf{S}$ 
is a good quantum number and the eigenvalue problem 
can be block-diagonalized
in terms of different seniority values. Loosely speaking, 
the seniority quantum number $\bf{S}$ is equal to the number of 
unpaired particles.
For systems with less than $\sim 16-18$ particles, 
this model can be diagonalized
exactly, and we can obtain {\em all eigenstates}. 
In our studies below, we will always
consider the case of half-filling, i.e., equally many particles
and single-particle levels. This case has the largest dimensionality: for 
16 particles in 16 doubly degenerate single-particle shells, we have a
total of $4\times 10^8$ states. We choose units MeV for the energy and 
set $G=0.2$ MeV in all calculations while we let $d$ vary. 

Through diagonalization 
of the above Hamiltonian
we can define exactly the
density of states $\Omega_N(E)$ for an $N$-particle
system with excitation energy $E$. 
An alternative 
to the exact diagonalization, would be 
to  use Richardson's well-known 
solution \cite{richardson}, 
however, we are interested in {\bf all} eigenstates,
and the amount of numerical labor will most likely be similar.
The density of states is
an essential ingredient in the
evaluation of thermal averages and for the discussion of phase
transitions in finite systems. For nuclei, experimental 
information on the density of states
is expected to reveal important information on nuclear shell
structure, pair correlations and other correlation phenomena
in the nucleonic motion. 

The density of states $\Omega_N(E)$ is the statistical
weight of the given state with excitation energy $E$, and its logarithm
\begin{equation}
      S_N(E)=k_B\ln\Omega_N(E),  \label{eq:entromicro}
\end{equation}
is the entropy (we set Boltzmann's constant
$k_B=1$) of the $N$-particle system.  
The density of states defines also the partition function in the 
microcanonical ensemble and can be used to compute the 
partition function $Z$ of the canonical ensemble through
\begin{equation}
    Z(\beta)=\sum_E\Omega_N(E)e^{-\beta E},
    \label{eq:canonicalpart}
\end{equation}
with $\beta=1/T$ 
the inverse temperature.  With $Z$ it is straightforward to 
generate other thermodynamical
properties such as the mean energy 
$\langle E\rangle$ or the specific heat $C_V$.

The density of states can also be used to define 
the  free energy $F(E)$ in the microcanonical ensemble 
at a fixed temperature $T$ (actually an expectation value in this ensemble), 
\begin{equation}
    F(E)=-T\ln\left[\Omega_N(E)e^{-\beta E}\right]\;.
    \label{eq:freenergy}
\end{equation}
Note that here we include only
configurations at a particular $E$.
 
The above free energy was used by e.g.,
Lee and Kosterlitz \cite{prl90},
based on the histogram approach for studying
phase transitions developed by Ferrenberg and Swendsen \cite{fs88},
in their studies of phase transitions
of classical spin systems. 
If a phase transition is present, a plot of $F(E)$ versus $E$ will show
two local minima which correspond to configurations that are
characteristic of the high and low temperature phases.
At the transition temperature $T_C$ the value of $F(E)$ at the 
two minima equal, while at temperatures below $T_C$, the low-energy
minimum is the absolute minimum. At temperatures above $T_C$, the high-energy
minimum is the largest. If there is no
phase transition, the system developes only one minimum for all temperatures.
Since we are dealing with finite systems, we can study the development 
of the two minima as function of the dimension of the system and thereby 
extract information about the nature of the phase transition. If we are dealing
with a second order phase transition, the behavior of $F(E)$ does not change
dramatically as the size of the system increases. However, if the transition
is first order, the difference in free energy, i.e., 
the distance between the maximum and minimum values, 
will increase with increasing dimension. 

\begin{figure}
\begin{center}
          {\epsfxsize=20pc \epsfbox{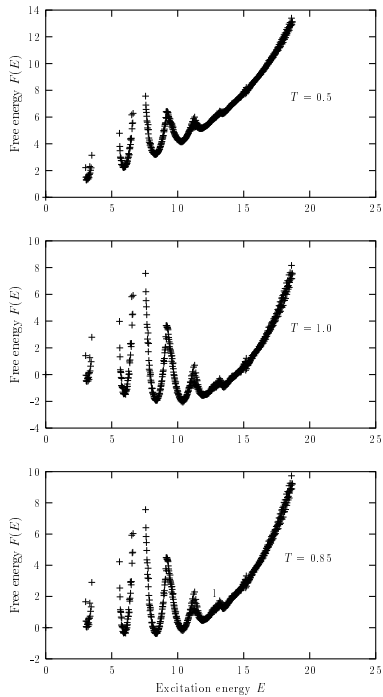}}
\caption{Free energy from Eq.~(\ref{eq:freenergy}) at $T=0.5$, $0.85$ and
         $T=1.0$ MeV  with 
         $d/G=0.5$ with 16 particles in 16 doubly degenerate
         levels. All energies are in units of MeV and 
         an energy bin of $10^{-3}$ MeV has been chosen.}
\label{fig:free_energy16}
\end{center} 
\end{figure}

To elucidate the nature of the transition
we calculate exactly the  
free energy $F(E)$ of Eq.~(\ref{eq:freenergy})
through diagonalization of the pairing Hamiltonian of 
Eq.~(\ref{eq:pairHamiltonian})
for systems with up to 16 particles in $16$ doubly degenerate
levels. 
For $d/G=0.5$ and 16 single-particle levels, 
we develop two clear  minima for the free energy.
This is seen in
Fig.~\ref{fig:free_energy16} where we show the free energy as function of 
excitation energy
using Eq.~(\ref{eq:freenergy}) at temperatures 
$T=0.5$, $T=0.85$ and $T=1.0$~MeV.
The first minimum corresponds to the case where we break one pair.
The second and third minima  correspond
to cases where two and three pairs are broken, respectively. 
When two pairs are broken, corresponding to seniority ${\bf S}=4$, 
the free energy minimum is made up of contributions
from states with ${\bf S}=0,2,4$. These contributions serve to lower
the free energy. 
Similarly, with three pairs
broken we see a new free energy minimum which receives contributions
from ${\bf S}=0,2,4,6$.
At higher excitation energies, population
inversion takes place, and our model is no longer realistic. 

We note that for $T=0.5$~MeV, the minima at lower excitation
energies are favored. 
At $T=1.0$~MeV, the higher energy
phase (more broken pairs) is favored.
We see also, at $T=0.85$~MeV, that 
the free-energy minima where we break two and three pairs 
equal. 
Where two minima coexist, we may have an
indication  of a phase transition. Note however that this is not a 
phase transition in the ordinary thermodynamical sense.
There is no abrupt transition from a purely paired phase to a 
nonpaired phase.  
Instead, our system developes several such intermediate steps
where different numbers of broken pairs can coexist. 
At e.g., $T=0.95$~MeV, we find again two equal minima. For this case,
seniority ${\bf S}=6$ and ${\bf S}=8$ yield two equal minima.
This picture repeats itself for higher seniority and higher temperatures.
\begin{table}[b]
\begin{center}
\begin{tabular}{ccccc}\hline
$N$ & 10 & 12 & 14& 16 \\\hline
$\Delta F/N$ [MeV]   &0.531 & 0.505 & 0.501 & 0.495 \\\hline
\end{tabular} 
\caption{ $\Delta F/N$ for $T=0.85$ MeV. See text for further details.} 
\end{center}
\label{tab:free_energy10_16}
\end{table}
If we then focus on the second and third minima, i.e., where we break
two and three pairs, respectively, the difference $\Delta F$ between the 
minimum and the maximum of the free energy, can aid us in distinguishing
between a first order and a second order phase transition. If $\Delta F/N$
remains constant as $N$
increases, we have a second order transition. An increasing $\Delta F/N$
indicates a first order phase transition. 
In Table \ref{tab:free_energy10_16} we display $\Delta F/N$ for 
$N=10$, 12, 14 and 16 at $T=0.85$ MeV. 
It is important to note that the features
seen in Fig.~\ref{fig:free_energy16}, apply to the cases with $N=10$, 12 
and 14 as well, where $T=0.85$~MeV is the temperature where the second and
third minima equal. This means that the temperature where the transition
is meant to take place remains stable as function of number of single-particle
levels and particles. This is in agreement with the simulations of 
Lee and Kosterlitz \cite{prl90}. We find a similar result for the minima
developed at $T=0.95$~MeV, where both ${\bf S}=6$ and ${\bf S}=8$ coexist.
However, due to population inversion, these minima are only seen clearly
for $N=12$, $14$ and $16$ particles.

Table \ref{tab:free_energy10_16} reveals that $\Delta F/N$ is nearly
constant, with  $\Delta F/N\approx 0.5$~MeV, indicating a 
transition of second order. This result is in 
agreement with what is expected for an infinite system. 
It is also easy to see from Fig.~\ref{fig:free_energy16},
that the entropy in the microcanonical ensemble can be convex for 
certain excitation energy ranges, resulting in eventual negative 
heat capacities, as inferred from the 
authors of Refs.~\cite{huller,gross}. The analysis
above however, does not lend
support to interpreting this as a sign of a first order phase transition.

We note the important result that for $d/G > 1.5$, 
our free energy, for $N\le 16$, developes
only one minimum for all temperatures. That is, for larger single-particle
spacings, there is no sign of a phase transition. This means that there
is a critical relation between $d$ and $G$ for the appearance of a phase 
transition-like behavior, being a  reminiscence of the thermodynamical limit.
This agrees also with e.g., the results for ultrasmall metallic grains
\cite{delft2000}. 

We have thus indications that the transition from the paired seniority
zero ground state to a mixed phase state is second order. The free-energy
analysis also demonstrates that each transition in seniority phases in 
the microcanonical ensemble is of second order. The strength of the
pairing in these systems determines the nature of the phase transitions. 
In particular, for a weakly paired system, we found no evidence 
for two phases, while normal pairing strengths, such as those
found in nuclei, may well exhibit the paired-phase and mixed seniority
phases that we demonstrated in this model. We will include more realistic
interactions to investigate this point in future work. 
We also found, using Auxiliary Field Monte Carlo 
computations for this system \cite{KD97} together 
with the histogram method of Refs.~\cite{prl90,fs88}, that the energy
fluctuations in the canonical ensemble 
make it rather difficult to extract useful information
on the nature of the phase transitions from these techniques. 

\section{Conclusions} \label{sec:sec5}
In summary, the $^1S_0$ and $^3P_2$ partial waves are crucial for our
understanding of superfluidity in neutron star matter. The role of polarization terms 
and hyperon pairing are
still open and unsettled topics, see ref.~\cite{dh2003} for further discussions.
Furthermore, we have also discussed 
recent experimental and theoretical studies 
of thermodynamical properties of finite nuclei
and their interpretation in terms  of eventual pairing transitions 
in finite nuclei. For a more detailed theoretical analysis we would howerver need 
extensive shell-model Monte Carlo simulations in order to test the
role played by e.g., pairing terms in the interaction. It is an open question whether
such calculations lend support to the experimentally observed   
level densities. 

\section*{Acknowledgements}
We are much indebted to 
Magne Guttormsen (Oslo)
and Andreas Schiller (LLNL/MSU) 
for the many discussion on the topics addressed here.

\end{document}